\documentclass[preprint1]{aastex}%
\usepackage{epsfig}
\makeindex 
\begin{document}
\title{ON THE INTERNAL DYNAMICS OF STARLESS CORES: STABILITY OF STARLESS CORES WITH INTERNAL MOTIONS AND COLLAPSE DYNAMICS}

\author{Young Min Seo\altaffilmark{1}, Seung Soo Hong\altaffilmark{2}, \& Yancy L. Shirley\altaffilmark{1,3}}
\affil{$^1$ Department of Astronomy \& Steward Observatory, University of Arizona, Tucson, AZ, USA}
\affil{$^2$ Astronomy Program, Department of Physics and Astronomy, Seoul National University, Seoul 151-742, Korea}
\affil{$^3$ Adjunct Astronomer at the National Radio Astronomy Observatory}

\begin{abstract}

In order to understand the collapse dynamics of observed low-mass starless cores, we revise the conventional stability condition of hydrostatic Bonnor-Ebert spheres to take internal motions into account. Because observed starless cores resemble Bonnor-Ebert density structures, the stability and dynamics of the starless cores are frequently analyzed by comparing to the conventional stability condition of a hydrostatic Bonnor-Ebert sphere. However, starless cores are not hydrostatic but have observed internal motions. In this study, we take gaseous spheres with a homologous internal velocity field and derive stability conditions of the spheres utilizing a virial analysis. We propose two limiting models of spontaneous gravitational collapse: the collapse of critical Bonnor-Ebert spheres and uniform density spheres. The collapse of these two limiting models are intended to provide the lower and the upper limits, respectively, of the infall speeds for a given density structure. The results of our study suggest that the stability condition sensitively depends on internal motions. A homologous inward motion with a transonic speed can reduce the critical size compared to the static Bonnor-Ebert sphere by more than a factor of two. As an application of the two limiting models of spontaneous gravitational collapse, we compare the density structures and infall speeds of the observed starless cores L63, L1544, L1689B, and L694-2 to the two limiting models. L1689B and L694-2 seem to have been perturbed to result in faster infall motions than for spontaneous gravitational collapse.

\end{abstract}

\section{INTRODUCTION}

Low-mass star formation is a fundamental process in astrophysics because the majority of stars in our Milky Way are low-mass stars. Recent observations at (sub)millimeter and IR wavelengths show that starless cores are the initial stage of low-mass star formation (Myers \& Benson 1983; Benson \& Myers 1989; Ward-Thompson et al. 1994). Isolated low-mass starless cores represent the simplest environment of star formation compared to high-mass star forming region (e.g. Orion, Cygnus, Carina, etc.) because they are in less crowded regions with less contamination from strong radiation fields of neighboring stars and from feedbacks of stellar outflows. A starless core usually forms a single star or a few stars at most (e.g. Lada 2006). Starless cores are therefore ideal objects to research the basic physics of star formation.

Star formation involves a series of complicated dynamical, chemical, and radiative processes. In one limit of theoretical studies, starless cores are formed through gravitational fragmentation and undergo hydrostatic collapse when they become gravitationally critical through loosing their thermal or magnetic pressure support (e.g. Mouschovias \& Ciolek 1999). In this limit, gravitational contraction is the dominant mechanism during collapse. Another limit of theoretical studies, which is still developing, proposes that turbulence is comparably important to self-gravity in forming stars  (e.g. Mac Low \& Klessen 2004; Kudoh \& Basu 2008; Nakamura \& Li 2008; Gong \& Ostriker 2009; Gong \& Ostriker 2011). In this limit, colliding flows in supersonic turbulence create or disrupt starless cores, and some fraction of starless cores made from colliding flows may undergo collapse. The main differences between this limit and the former one are that the speed of colliding flows may be much faster than the speed of gravitational contraction and that the ram pressure of colliding flows may considerably shorten the time to form stars. Thus, studies of the internal velocity fields of starless cores are necessary to determine which limit is dominant.

The internal velocity fields of starless cores are probed using dense gas tracers such as HCO$^+$, HCN, CS, etc. (e.g Lee et al. 1999, 2004; Gregerson et al. 2000; Evans et al. 2005; Sohn et al. 2007; Stahler \& Yen 2009).  Transitions of these tracers are usually optically thick lines which have an asymmetric profile compared to optically thin symmetric lines when there are internal motions. Surveys of the skewness of CS 2$-$1, CS 3$-$2, and HCN 1$-$0 lines were done by Lee et al. (1999, 2004), and Sohn et al. (2007). Their studies show that 25\%, 27\%, and 20\% of their samples, respectively, have blue-skewed line profiles indicative  of infall. In addition, studies on the internal velocity fields of L694-2 and L1197 with radiative transfer models and dynamical simulations (Lee et al. 2007; Seo et al. 2011b) suggest that both starless cores have supersonic infall motions substantially faster and occurring in a narrower layer than predicted by quasi-static gravitational collapse. Such fast infall motions may appear in a scenario of colliding shocks where the turbulent surrounding may enhance infall speed in a narrow layer (G\'{o}mez et al. 2007) and considerably affect stability of the starless cores because the ram pressure $P_{\rm ram} \sim \rho v^2$ becomes comparable to the surface pressure $P_{\rm surf} \sim \rho c_s^2$. These studies demonstrate that internal motions of some starless cores may not be owing completely to gravitational contraction, but may also be under considerable influences from their surrounding environments. Their internal motions must be taken into account when testing their stability and understanding their internal dynamics.

Dynamical evolution of starless cores in the limit of negligible external dynamics (gravity dominant limit) are relatively well studied utilizing similarity solutions and numerical simulations (e.g. Larson 1969; Penston 1969; Shu 1977; Hunter 1977; Foster \& Chevalier 1993; Aikawa et al 2005; Seo et al. 2011b). The fiducial model in this limit is the collapse of a self-gravitating, isothermal, hydrostatic gas sphere bound by external pressure called the Bonnor-Ebert sphere (Ebert 1955; Bonnor 1956; Ebert 1957; hereafter BE).  This model has been widely used because it is a physically well established model with its column density profile similar to observed column densities in near-IR absorptions and dust continuum emissions (e.g., Alves et al. 2001; Evans et al. 2001; Tafalla et al. 2002; Harvey et al. 2003a, 2003b; Kandori et al. 2005). In this model, the critical BE sphere collapses while maintaining a BE-like density profile (Foster \& Chevalier 1993; Aikawa et al 2005; Seo et al. 2011b). However, the initial BE sphere is a hydrostatic sphere neglecting any internal motion, whereas real starless cores are likely contracting, expanding, or oscillating (e.g. Lee et al 1999, 2004; Keto et al. 2006).  The dynamics of starless cores in the limit of external dynamical influences (turbulence regulating limit) are also studied through simulations (e.g. Kudoh \& Basu 2008; Nakamura \& Li 2008), but do not have a simple fiducial model because of the chaotic nature of turbulence. Since only the first limit has a fiducial model, it is hard to compare observed starless cores to the simulations and interpret their dynamics with respect to the two limits. A dynamics model is required that provides a boundary between the two limits.

The main purpose of this work is to take an internal velocity field into account in the determination of the dynamical stability condition for starless cores. Our stability condition serves as a more realistic stability criterion for starless cores with internal motions. We also suggest a range of infall velocities at a given density structure that may classify observed starless cores with respect to the two dynamical evolution limits. The comparison of observed infall speeds of starless cores to this range will tell us whether the collapse of starless cores mostly depend on its internal properties or evolve together with its turbulent surroundings.

In this paper, first, we derive a new dynamical stability condition of the BE spheres with various dimensionless size and homologous infall motions. Hunter (1979) studies stability of a uniform field with homologous infall motions using the virial treatment. We apply Hunter's analysis to the BE spheres and derive critical sizes of BE spheres for a given homologous infall motion. We define $\xi_{\rm max}$ as the dimensionless size
\begin{eqnarray}
\xi_{\rm max} \equiv { R_s(t) \sqrt{4 \pi G \rho_c(t)} \over{c_s} },
\label{xi_max}
\end{eqnarray}
where $t$ is time, and $G$ is the gravitational constant, $R_s$ and $\rho_c$ are the outer radius and the central density of the core. We calculate dynamical domains in a two dimensional $\xi_{\rm max}$ $vs.$ peak infall velocity space. The critical size $\xi_{\rm crit}$ for a given infall speed will separate the stable and unstable regimes in the $\xi_{\rm max}$ {\it vs.} peak infall velocity space.

Second, we search a range of infall velocities for isothermal spheres without any initial internal motion and any external perturbation. This is to find dynamical domain of the pure gravitational collapse in the $\xi_{\rm max}$ {\it vs.} peak infall velocity space. Without any initial internal motion and external perturbation, the dynamics of the starless core depend only on the density distribution and there are two limits of the density distributions for starless cores: the BE sphere and a uniform sphere. The collapse of the critical BE sphere gives the slowest infall with respect to the density concentration (Seo et al. 2011b) and provides the lower limit of the infall speed in spontaneous gravitational collapse (or pure gravitational collapse). A starless core that collapses slower than this model is likely to be perturbed by external perturbations or is not isothermal. The uniform sphere is a starless core with the least pressure gradient against the self-gravity for its initial conditions. Inevitably, the sphere develops a faster infall motion than the collapse of the critical BE sphere at a given density structure (Seo et al. 2011b). Since the uniform density field has the shallowest profile that a starless core can have, the collapse of the critical uniform sphere has the fastest infall speed with respect to a given density concentration (in the absence of external perturbations). Any starless cores collapsing faster than the collapse of the critical uniform sphere is likely to have initial converging flows, or be perturbed to collapse faster than any pure gravitational collapse by an external source.

As an application of our study, four starless cores (L1544, L1689B, L63, and L694-2) are compared with the new stability condition and the domains of two dynamical limits (gravity dominant and turbulence regulating limits). The four starless cores are the best four infall candidates with studies of both density structures and internal motions in dust continuum or extinction and molecular tracers of infall (e.g. HCN, CS, and HCO$^+$). In order to normalize the physical quantities of the four cores into dimensionless quantities of this study, we need to measure gas temperatures of the starless cores. We observed the four starless cores in NH$_3$ (1,1) and (2,2) inversion lines with the Robert C. Byrd Green Bank Telescope and deduced their gas kinetic temperature. Their collapse dynamics are interpreted with the results of our dynamical calculations.

The layout of this paper is as follows: \S2 briefly introduces our observation of NH$_3$ lines measuring temperatures of the four starless cores. In \S3 we introduce the result of Hunter (1979) and elaborate our application of his method to an arbitrary density distribution. In \S4 we outline the two limiting models of the spontaneous gravitational collapse, and discuss the physical meaning of the regimes divided by the two models in the $\xi_{\rm max}$ {\it vs.} peak infall speed space. In \S5 we discuss limitations of our analysis and results. We also compare observations of the four starless cores with our stability conditions and discuss their internal dynamics. Finally, in \S6, we will summarize our results and discussion.

\section{GBT OBSERVATIONS}
Observations of the NH$_3$ (1,1) and (2,2) inversion transitions were performed over three observing shifts in 2006 (GBT06A68) with the Robert C. Byrd Green Bank Telescope\footnote{The National Radio Astronomy Observatory is a facility of the National Science Foundation operated under cooperative agreement by Associated Universities, Inc.}. Each observations consisted of a 4 minute frequency switched observation with a 4.11 MHz throw at 4 Hz centered on the 850 $\mu$m continuum peak position (see Table 1).  The GBT spectrometer was set up with 50 MHz bandwidth and 6.1 kHz spectral resolution (0.08 km/s resolution).  The spectra were folded and baselined using standard GBTIDL routines.  The main beam efficiency was determined from observations of quasars (3C48 and 3C286) and planets to be $\eta_{mb} = 0.74 \pm 0.05$. In Section 5 we analyze the NH$_3$ observations for the four starless cores with measured infall motions (L1544, L1689B, L63, and L694-2).

\section{GRAVITATIONALLY STABLE OR UNSTABLE: VIRIAL TREATMENT}

To facilitate the virial analysis for a starless core, we assume the core is an isothermal gas sphere bounded by external pressure. We also assume that all internal motions are spherically symmetric. The virial theorem for an isothermal medium with spherical symmetry is given by
\begin{eqnarray}
{1 \over 2}{d^2 \over dt^2} \int_0^{R_{\rm{s}}}\rho(r)r^2\cdot4\pi r^2dr = \int_0^{R_{\rm{s}}}4\pi r^2 \rho(r)\dot{r}^2 dr + 3\int_0^{R_{\rm{s}}}4\pi r^2\rho(r)c_s^2 dr \nonumber
\\ -\int_0^{R_{\rm s}}4\pi r^2\rho(r){G\int_0^r 4\pi r'^2\rho(r') dr'\over r}dr -4\pi R_s^3c_s^2\rho(R_s),
\label{virial_sphere}
\end{eqnarray}
where $R_s$ is the radius of the core, and $c_s$ is the sound speed. The above equation may be normalized with the following normalizing parameters:
The sound speed at the core is
\begin{eqnarray}
c_{s,c}~=~0.188~{\rm km~s^{-1}}~\left({T \over 10~{\rm K}}\right)^{1/2} + \sigma_{\rm nt},
\label{csc}
\end{eqnarray}
where $T$ means the temperature of the core, and $\sigma_{\rm nt}$ is the {\it rms} speed of non-thermal components. The normalizing parameter for length $\alpha$ is defined as
\begin{eqnarray}
\alpha~\equiv~{c_s\over \sqrt{4\pi G \rho_c} }=~0.044~{\rm pc}~\left({c_s\over 0.188~{\rm km~s^{-1}}}\right)~\left({10^4~{\rm cm^{-3}} \over n_c}\right)^{1/2}~\left({2.33 \over \mu}\right)^{1/2},
\label{alpha}
\end{eqnarray}
where $n_c$ is the number density of molecules at the core center, and $\mu$ is the mean molecular weight. For the normalization of time, we take the sound crossing time $t_0$, which is the time taken by the sound wave to travel a unit length,
\begin{eqnarray}
t_0~\equiv~{1\over \sqrt{4\pi G \rho_c} }~=~ 0.23 ~ \times ~10^6 ~ {\rm yr} ~ \left({10^4 ~ {\rm cm^{-3}} \over n_c}\right)^{1/2}~\left({2.33 \over \mu}\right)^{1/2}.
\label{tau}
\end{eqnarray}
The normalized form of equation (\ref{virial_sphere}) is
\begin{eqnarray}
\nonumber
{d^2\over d\tau^2} \int_0^{\xi_{\rm s}}s(\xi)\xi^4d\xi = 2\int_0^{\xi_{\rm s}}s(\xi)\dot{\xi}^2\xi^2d\xi + 6\int_0^{\xi_{\rm s}}s(\xi)\xi^2d\xi -{2\over3}\int_0^{\xi_{\rm s }}\xi s(\xi)\tilde{M}(\xi)d\xi \\-2s(\xi_{\rm s})\xi_{\rm s}^3,
\label{norm_virial}
\end{eqnarray}
where $s\equiv \rho/\rho_c(t=0)$, $\xi \equiv r/\alpha$, $\xi_s \equiv R_s/\alpha$, $\tilde{M}(\xi)=\int_0^\xi\xi^2 s(\xi) d\xi$ and $\tau\equiv t/t_0$.

\subsection{A Uniform Density Core with a Homologous Motion}

The dynamical stability condition of a uniform density field with homologous infall motions is derived by Hunter (1979). He calculated how fast the converging motion should be to make a stable core into a critical configuration after contraction. In this paper, we show his results and briefly discuss its meaning in $\xi_{\rm max}$ {\it vs.} peak infall speed space. For detailed derivation, please refer to his paper.

The relation between the initial speed of homologous motion and the ratio of initial and final size of the core (equation (4) in Hunter's paper) in terms of our dimensional variables is as follow:
\begin{eqnarray}
U_{\rm i }^2=10\left[\ln\mathbf{R}-\left({4\over3}-{1\over\mathbf{R}}\right)\left(1-{1\over\mathbf{R}^3}\right)\right],
\label{v5}
\end{eqnarray}
where $U_i$ is the initial peak velocity of homologous motion divided by the sound speed. $\mathbf{R}$ is the compression ratio defined as $\xi_{\rm si}/\xi_{\rm sf}$ where $\xi_{\rm si}$ is the core radius at the initial moment and $\xi_{\rm sf}$ is the core radius at the final moment. When there is no external pressure, equation (\ref{v5}) becomes
\begin{eqnarray}
U_{\rm i }^2=10\left(\ln\mathbf{R}+{1\over\mathbf{R}}-1\right).
\label{v6}
\end{eqnarray}

\begin{figure}[tbh]
  \begin{center}
    \leavevmode
    \epsfxsize = 13.5cm
    \epsfysize = 7.cm
    \epsffile{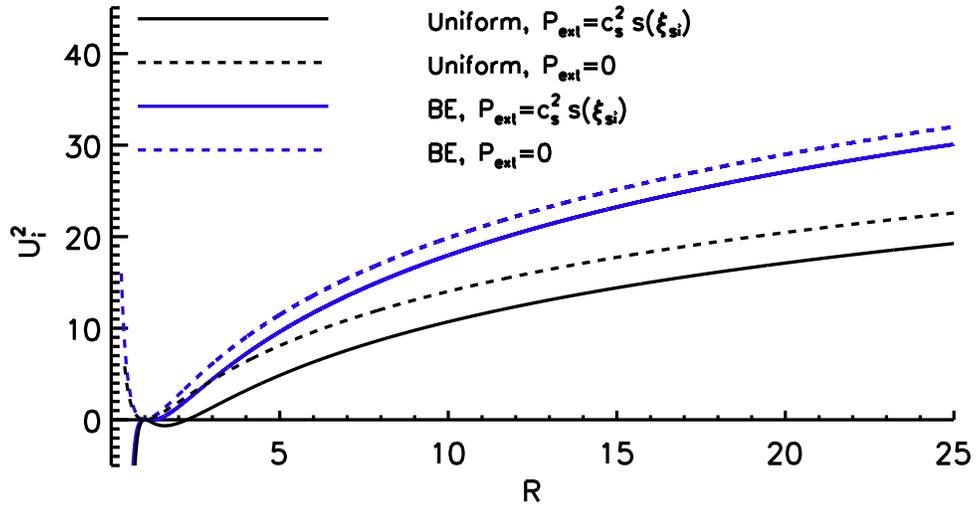}
  \end{center}
\caption{{ \sl The relationship between the inward homologous motion and the compression ratio for a uniform density core (black) and a Bonnor-Ebert core (blue). The pressure confined core (solid) is easier to compress than the core in a vacuum (dashed). Stronger compression occurs by a faster inward motion.
}}
\label{fig1}
\end{figure}

The initial velocity $U_{\rm i}$ are plotted in black lines as functions of the compression ratio $\mathbf{R}$ in Figure \ref{fig1}. The solution of $U_{\rm i}^2 < 0$ is not a true solution because the velocity becomes imaginary. The solution of $U_{\rm i}^2>0$ and $\mathbf{R}>1$ is for a collapsing core, while the one of $\mathbf{R}<1$ is for a expanding core. The solutions for the cores bound with the external pressure of $P_{\rm ext}=c_s^2s(\xi_{\rm si})$ and $P_{\rm ext}=0$ are marked with the solid and dashed lines, respectively. The solution of $P_{\rm ext}=c_s^2s(\xi_{\rm si})$ has imaginary values from $\mathbf{R}$ = 1 to 2.25. This is because we assume that the density structure is kept as a uniform field during the contraction, which is unrealistic. Because of this assumption, the work done by the external pressure during the contraction cannot be correctly estimated. On the other hand, the solution without the external pressure has no imaginary values since the stability depends only on the gravitational potential energy and the kinetic energy.

\begin{figure}[tbh]
  \begin{center}
    \leavevmode
    \epsfxsize = 13.5cm
    \epsfysize = 11.cm
    \epsffile{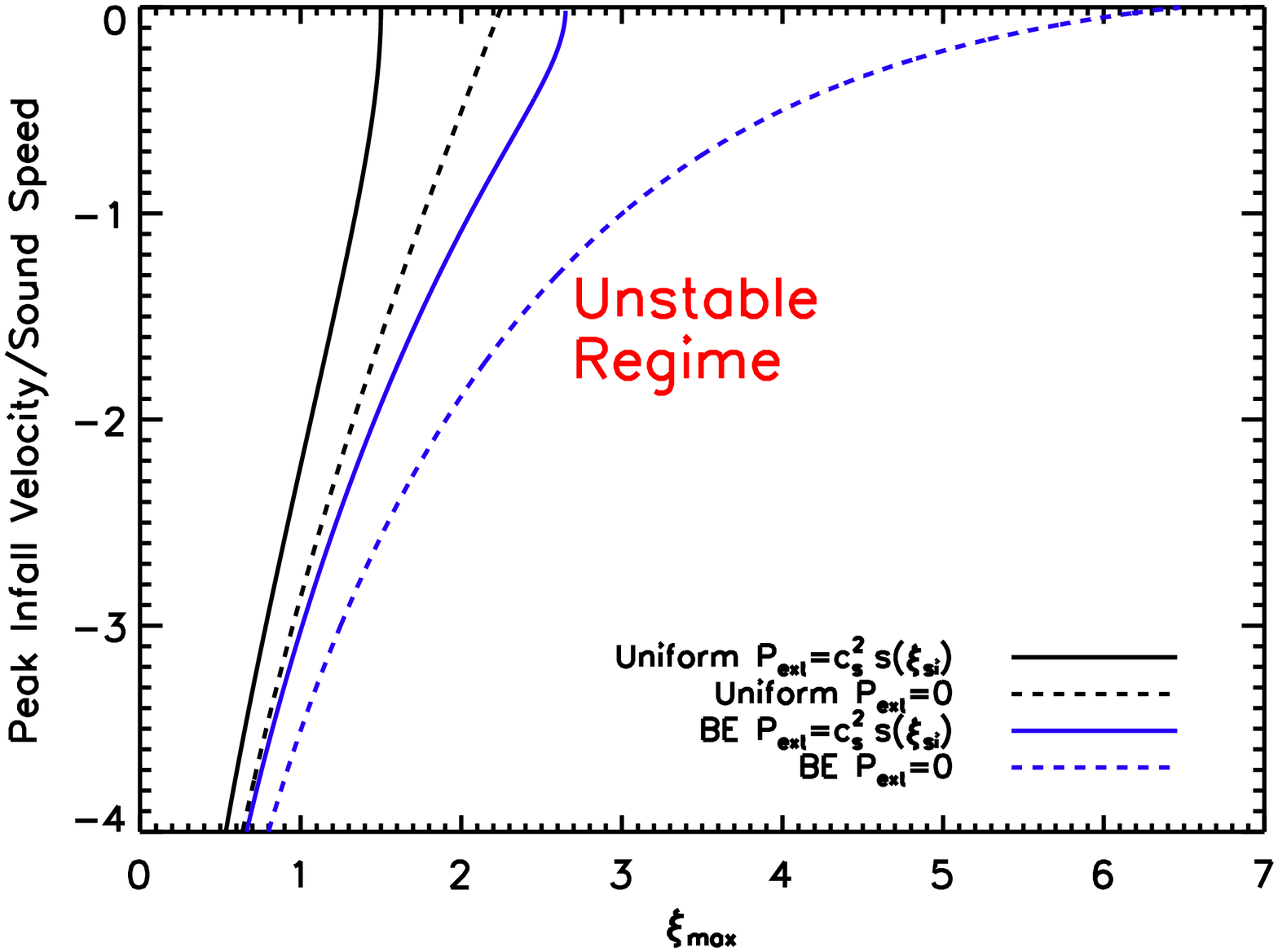}
  \end{center}
\caption{{ \sl Critical size of the uniform density core (black) and the BE core (blue) with respect to the peak speed of inward motion.  The solid and dashed lines denote cores bounded by two limits of the external pressure.}}
\label{fig2}
\end{figure}

Since core mass is fixed and we assume the uniform density structure is maintained, the radius and the central density of the core after contraction are $R_{\rm s}/\mathbf{R}$ and $\rho_c(t=t_{\rm f}) = \mathbf{R}^3\rho_c(t=0)$, where $t_{\rm f}$ is the time epoch at the completion of contraction by inward motions. So, the dimensionless size $\xi_{\rm max}$ becomes $\xi_{\rm max}(t=t_{\rm f}) = \xi_{\rm max}(t=0)\mathbf{R}^{1/2}$. The final dimensionless size $\xi_{\rm max}(t=t_{\rm f})$ of the core should be the same as the critical size of the static uniform density core. The critical size of the pressure-bound, static uniform density core is $\xi_{\rm max}=$2.25, which is numerically estimated using a Godunov-type hydrodynamics code within an error of 5\% (Seo et al. 2011a, 2011b). For a given inward velocity $U_{\rm i}$ and the final critical size $\xi_{\rm max}(t=t_{\rm f})$, the compression ratio $\mathbf{R}$ is determined, and we can derive the corresponding $\xi_{\rm max}(t=0)$ using equations (\ref{v5}) and (\ref{v6}). The stability diagram for a uniform density cores with inward homologous motions is presented in Figure \ref{fig2}. The black solid line represents the core confined by the external pressure, and the black dashed line represents core with no external pressure. The regime at the left side of the black dashed line is where a core becomes gravitationally stable, while the one at the right side of the black solid line is the unstable domain for the critical uniform density core. The stability in the regime between the two lines depends on the external pressure.

\subsection{The Bonnor-Ebert Sphere with a Homologous Motion}

Observed starless cores are not a uniform density cores, but are reported to have BE-like density profiles.  If the density field is not constant, but a function of the radial distance, the moment of inertia $I$ and the gravitational energy $W$ take forms of $C_0(\xi_{\rm s})\tilde{M}(\xi_{\rm s})\xi_{\rm s}^2$ and $C_1(\xi_{\rm s})\tilde{M}(\xi_{\rm s})/\xi_{\rm s}$, respectively, where $C_0(\xi_{\rm s})$ and $C_1(\xi_{\rm s})$ are functions of the radial distance. If the $C_0$ and $C_1$ do not sensitively depend on the radial distance and time during collapse, we may assume them as constants. The $C_0$ value decreases from $0.2$ to $0.14$ for $\xi_{\rm max}$ from 0 to 7 and becomes almost constant for $\xi_{\rm max}\geq$10. Since the total variation of $C_0$ is about 30\% for smaller cores of $\xi_{\rm max}<6.5$ while the observational uncertainty in $\xi_{\rm max}$ is at least a factor of two (Kandori et al. 2005), then assuming $C_0$ as a constant is a reasonable assumption unless the core is considerably disturbed and deviates from a BE-like structure (quasi-equilibrium contraction). Likewise, the $C_1$ value varies from 0.6 to 1 (40\%), and thus we may also assume it is approximately a constant. In this study, we take $C_0$ and $C_1$ as $C_0 = C_0(\xi_{\rm si})$ and $C_1 = C_1(\xi_{\rm si})$, respectively.

The virial theorem for the BE sphere is given by
\begin{eqnarray}
{d^2\over d\tau^2}C_0\tilde{M}(\xi_{\rm s})\xi_{\rm s}^2=2\dot{\xi_{\rm s}}^2C_0\tilde{M}(\xi_{\rm s})+2\tilde{M}(\xi_{\rm s})-2C_1{\tilde{M}^2(\xi_{\rm s})\over\xi_{\rm s}}-2\xi_{\rm s}^3s(\xi_{\rm s}).
\label{be_v1}
\end{eqnarray}
We assume that the core does not get any additional mass during contraction so the core mass $\tilde{M}_\circ$ is a constant $\tilde{M}_\circ=\tilde{M}(\xi_{\rm si})$. Assuming that the inward motion is a homologous motion, we have
\begin{eqnarray}
\ddot{\xi_{\rm s}}={1\over C_0}{1\over\xi_{\rm s}}-{C_1\over C_0}{\tilde{M}_\circ\over\xi_{\rm s}^2}-{\xi_{\rm s}^2\over C_0}{s(\xi_{\rm s})\over \tilde{M}_\circ}.
\label{be_v2}
\end{eqnarray}
We integrate the above equation from $\xi_{\rm{sf}}$ to $\xi_{\rm{si}}$ and assume that the external pressure is constant $s(\xi_{\rm si})=s(\xi_{\rm sf})$. Then,
\begin{eqnarray}
U_{\rm i}^2=U_{\rm f}^2+{2\over C_0}\ln({\xi_{\rm si}\over\xi_{\rm sf}})+2{C_1\over C_0}\tilde{M}_\circ\left({1\over\xi_{\rm si}}-{1\over\xi_{\rm sf}}\right)-{2\over3C_0}{s(\xi_{\rm si})\over \tilde{M}_\circ}(\xi_{\rm si}^3-\xi_{\rm sf}^3),
\label{be_v3}
\end{eqnarray}
where $U_{\rm f}$ is the velocity of the homologous motion at the final moment. Since the total mass of the core is fixed, we may write the core mass $\tilde{M}_\circ=C_2(\xi_{\rm si})\xi_{\rm si}^3$. The equation (\ref{be_v3}) becomes
\begin{eqnarray}
U_{\rm i}^2=U_{\rm f}^2+{2\over C_0}\ln{\xi_{\rm si}\over\xi_{\rm sf}}+2{C_1\over C_0}{\tilde{M}_\circ\over\xi_{\rm sf}}\left({\xi_{\rm sf}\over\xi_{\rm si}}-1\right)-{2\over3C_0C_2}s(\xi_{\rm si})\left(1-{\xi_{\rm sf}^3\over \xi_{\rm si}^3}\right).
\label{be_v4}
\end{eqnarray}
We assume that the kinetic energy of the homologous motion is expended at the final moment ($U_{\rm f}=0$) and the core becomes relaxed ($\ddot{\xi_{\rm sf}}=0$). Equation (\ref{be_v2}) at the final moment is
\begin{eqnarray}
{1\over C_0}-{s(\xi_{\rm si})\over C_0C_2}{\xi_{\rm sf}^3\over\xi_{\rm si}^3}={C_1\over C_0}{\tilde{M}_\circ\over\xi_{\rm sf}}
\label{be_v5}
\end{eqnarray}
Putting the above result into the equation (\ref{be_v4}), we get
\begin{eqnarray}
U_{\rm i}^2={2\over C_0}\left[\ln\mathbf{R}+\left(1-{s(\xi_{\rm si})\over C_2} {1\over\mathbf{R}^{3}}\right)\left({1\over\mathbf{R}}-1\right)-{s(\xi_{\rm si})\over3C_2}\left(1- {1\over\mathbf{R}^3}\right)\right].
\label{be_v6}
\end{eqnarray}
When there is no external pressure, the above equation becomes
\begin{eqnarray}
U_{\rm i}^2={2\over C_0}\left(\ln\mathbf{R}+{1\over \mathbf{R}}-1\right).
\label{be_v7}
\end{eqnarray}

In Figure \ref{fig1} the square of initial velocity of equations (\ref{be_v6}) and (\ref{be_v7}) are plotted in blue lines as a function of the compression ratio $\mathbf{R}$. The solid blue line represents the solution of $P_{\rm ext}=c_s^2s(\xi_{\rm si})$, and the dashed blue line represents the solution without the external pressure. Only the solutions at $U_{\rm i}^2>0$ and $\mathbf{R}>1$ correspond to collapsing cores. In this figure, the $C_0$ is fixed to be $C_0(\xi_{\rm max}=6.5)$. The solution of $P_{\rm ext}=c_s^2s(\xi_{\rm si})$ also has imaginary velocity at the range of $\mathbf{R}$ from 1 to 1.5 because the assumption of keeping a BE-like structure during the contraction results in incorrectly estimating the work done by the external pressure.

The critical size of BE cores with inward homologous motions can be calculated as follows:
The final radius of the core after contraction by the inward motion, $R_{\rm sf}$, is given by
\begin{eqnarray}
R_{\rm sf}=\alpha\xi_{\rm sf}=\alpha_{\rm f}\xi_{\rm max,f},
\label{rsf}
\end{eqnarray}
where $\alpha_{\rm f}$ and $\xi_{\rm max,f}$ are the newly defined normalizing parameter and dimensionless size of the core, respectively, after the contraction. Since $\alpha\sim \rho_c^{-1/2}$, the equation (\ref{rsf}) may be reduced to
\begin{eqnarray}
\xi_{\rm max,f} = \left({\rho_{\rm cf} \over \rho_{\rm ci}}\right)^{1\over2} {\xi_{\rm si}\over \mathbf{R}},
\label{xif}
\end{eqnarray}
where the $\rho_{\rm ci}$ and $\rho_{\rm cf}$ are the central densities at the initial moment and after the contraction, respectively. The relation between the central densities is given by the conservation of the total mass of the core,
\begin{eqnarray}
4\pi \alpha_i^3\xi_{\rm si}^2\left.{d\psi\over d\xi}\right|_{\xi_{\rm si}}= 4\pi \alpha_f^3\xi_{\rm max,f}^2\left.{d\psi\over d\xi}\right|_{\xi_{\rm max,f}},
\label{tm}
\end{eqnarray}
where $\psi\equiv -\ln s$. Reducing the above equation and combining with the equation (\ref{xif}), we have
\begin{eqnarray}
\mathbf{R}={\xi_{\rm max,f}\left.{d\psi\over d\xi}\right|_{\xi_{\rm max,f}} \over \xi_{\rm si}\left.{d\psi\over d\xi}\right|_{\xi_{\rm si}}}.
\label{rBE}
\end{eqnarray}
To be a critical core, the final size of the core $\xi_{\rm max,f}$ should be equal to the critical size of the static BE sphere, $\xi_{\rm max,f}=6.5$. If the initial size is given, the compression ratio $\mathbf{R}$ can be estimated for a critical core using the above equation. The corresponding peak velocity of the homologous motion can be calculated from the equation (\ref{be_v7}). Thus, we may relate an internal motion that makes a core to be critical with an initial size of a BE sphere.

The critical sizes of BE cores with homologous internal motions are plotted in Figure \ref{fig2}. The blue solid and dashed lines are the critical sizes of the BE cores with the external pressure of $P_{\rm ext}=c_s^2s(\xi_{\rm si})$ and without external pressure, respectively. Because of the imaginary velocity in the solution of $P_{\rm ext}=c_s^2s(\xi_{\rm si})$, the $\xi_{\rm max}$ value of the blue solid line starts from much smaller values than the critical size of the static BE core. The choice of $C_0$ and $C_1$ changes the starting points of lines for pressure bound cores, but not by more than factor of two. The critical sizes of smaller BE cores converge to those of uniform density cores because smaller BE cores have flatter density profiles. The critical size is very sensitive to internal motions. A homologous inward motion with a transonic speed can reduce the critical sizes of uniform density cores and BE spheres by more than half.

\section{SPONTANEOUS COLLAPSE OF STARLESS CORES AND LOWER AND UPPER LIMITS OF INFALL SPEEDS}

The collapse of the critical BE spheres and uniform density spheres provide the lower and upper limits, respectively, of infall speeds in the spontaneous gravitational collapse of isothermal cores. Figures 3 and 4 of Seo et al. (2011b) show density and velocity distributions of the collapse of the critical BE and uniform density spheres, respectively. Their density structures resemble each other at the later stage of evolution, but infall motions of the collapse of the critical uniform density sphere is always faster than the collapse of the critical BE sphere.

\begin{figure}[tbh]
  \begin{center}
    \leavevmode
    \epsfxsize = 13.5cm
    \epsfysize = 11.cm
    \epsffile{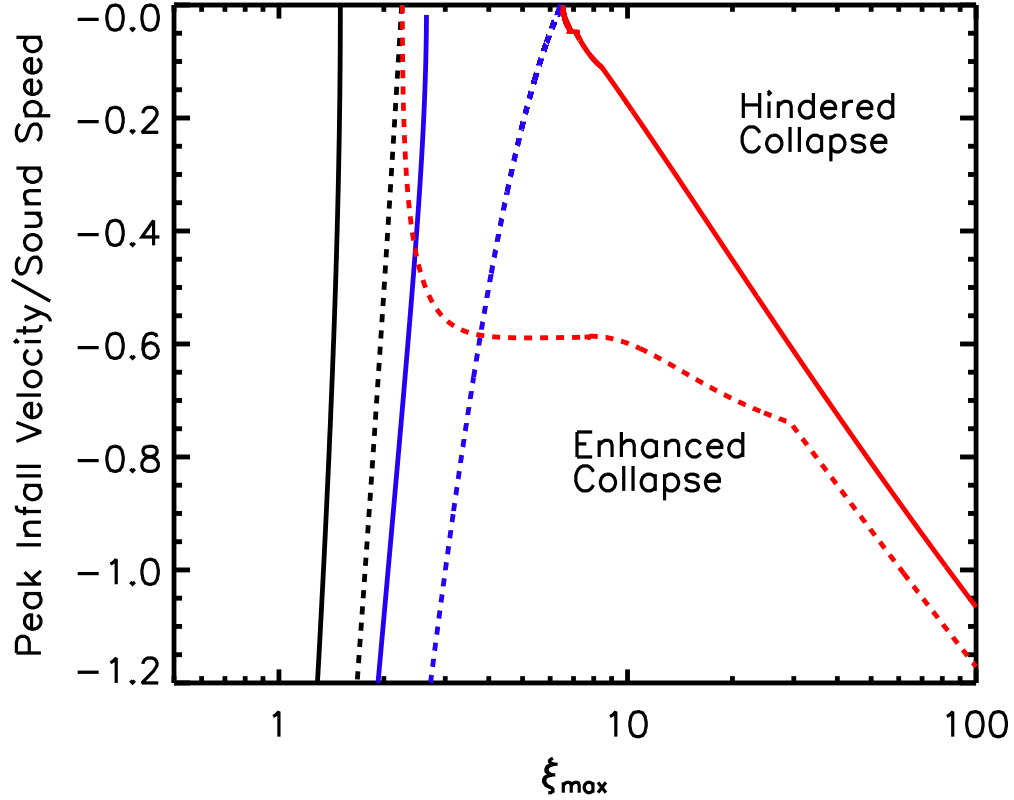}
  \end{center}
\caption{{ \sl The evolution of the marginally unstable uniform density (red dashed) and BE (red solid) cores, and the critical sizes of the uniform density (black) and BE (blue) cores with respect to the peak speed of inward motion. The collapse of the BE core delivers the slowest infall motions, while the collapse of the uniform density core results in the fastest gravitational collapse. If a starless core goes through purely gravitational collapse, it is likely to be between the two evolution lines. If collapse of starless core is hindered by external perturbations, it may collapse slower than the collapse of the BE core. On the other hand, a starless core collapse faster than the collapse of the uniform density core is likely to be perturbed by external perturbations to enhance its infall speed.
}}
\label{fig3}
\end{figure}

From the numerical simulations of the collapse of the critical BE and uniform density spheres, we measure the peak infall velocities of the two models as a function of $\xi_{\rm max}$. The results are plotted in Figure \ref{fig3}. The red solid line represents the collapse of the critical BE sphere, and the red dashed line represents the collapse of the critical uniform density sphere. The stability lines are also plotted in the same figure with the same colors and line styles used in Figure \ref{fig2}. The collapse of the critical BE sphere shows a steady and slow growth of infall speed along with $\xi_{\rm max}$. On the other hand, the collapse of the critical uniform density sphere results in a sudden increase of infall speed without a considerable density concentration because there is no pressure gradient at the initial moment. Owing to the sudden infall motion at the early evolution stage, the core bounces momentarily and generates a strong accretion shock, which travels outward and regulates the growth rate of infall motion. The accretion shock is shown as a kink or cusp in the velocity fields in Figures 3 and 4 of Seo et al. (2011b). The kinks in Figure \ref{fig3} at $\xi_{\rm max}= 8$ and 30 are when the accretion shock passes through the peak infall layer and when it escapes the core, respectively.

Since the collapse of the critical BE sphere provides the lower infall speed limit of the spontaneous gravitational collapse, the regime above the red solid line implies that the gravitational collapse is hindered by an external perturbation. The regime below the collapse of the critical uniform density sphere is for gravitational collapse enhanced by an external perturbation. For a starless core that evolves between the two lines, it is hard to tell which factors, either initial density distribution or external perturbation or both, determines its dynamical evolution. Both factors may be equally responsible.

\section{DISCUSSIONS}

\subsection{Limitations of Our Analysis}

The stability of isothermal gas spheres with homologous motions is derived utilizing the virial theorem. We impose homologous internal motions because they facilitate an analytic analysis. However, there is no observational evidence that starless cores have homologous internal motions when they are collapsing or perturbed. Moreover, a starless core does not maintain a homologous velocity field during its collapse even when a homologous velocity field is given as an initial perturbation. Nevertheless, our virial analysis provides a rough estimate of how much the conventional stability condition of BE sphere varies when there are inward velocity fields. Results in this paper suggest that the stability condition for starless cores is very sensitive to internal motions. A transonic homologous inward motion can reduce the critical size of the static uniform density and BE cores by about 67\% and 54\%, respectively.

Spherically symmetric starless cores are assumed in this study. For a spheroidal core in a hydrostatic state, its critical size is the same as the critical size of the BE sphere as long as its projected area is the same as that of the BE sphere (Lombardi \& Bertin 2001). Since the virial theorem is essentially an energy analysis, a stability analysis for spheroidal cores would be the same as this study except for the values of $C_0$ and $C_1$. Infall motions of the non-spherical cores were studied by Myers (2005) whom showed that the peak infall speeds are about 50\% faster for a cylindrical core and about twice faster for an infinitely extended slab. Observed starless cores usually have $\leq$2:1 axis ratios (Myers et al. 1996), which is geometrically closer to the sphere than the cylinder. So, the evolution lines of the two models in Figure \ref{fig3} may have slightly faster infall speeds if the cores are spheroids, but the difference should be quite a bit less than 50\% of the plotted value.

Isothermality is another assumption in this study. Starless cores are observed to have temperature profiles decreasing inward because the interstellar radiation field heats the outskirts of the cores (e.g. Evans et al. 2001; Zucconi et al. 2001; Ward-Thompson et al. 2002; Pagani et al. 2004; Shirley et al. 2005; Crapsi et al. 2007; Launhardt et al. 2013). We assumed isothermality because it simplifies the dynamics calculation by obviating the need to perform radiative transfer calculations. The effects due to non-isothermal temperature profile have been studied by Sipil\"{a}, Harju, \& Juvela (2011). They demonstrated that the critical sizes of BE-like cores depend on their temperature profiles; however, the critical size varies only about 6\% (from 6.38 to 6.76) from the original critical size (6.5), while the temperature varies 35\% (from 7 K to 9.5 K). Considering that the observed measurements of $\xi_{\rm max}$ has a typical error of 50\%, our results with an isothermal assumption are still applicable to real starless cores.

Our analysis also ignores rotation of starless cores. Studies of molecular line widths toward starless cores show that cores slowly rotate (e.g. Heiles \& Katz 1976; Arquilla \& Goldsmith 1986; Goodman et al. 1993; Caselli et al. 2002). If the rotational energy is comparable to the gravitational energy, rotation cannot be ignored in studying dynamical stability of starless cores. The typical observed ratio of the rotational energy to the gravitational energy is about 0.02 (Goodman et al. 1993). So, our assumption of ignoring rotation is practical for typical starless cores.

Magnetic fields may support starless core against gravitational collapse. Magnetic field strength of starless cores from Zeeman splitting measurement and from polarization observations range from 10 $\mu$G to $\leq$ 200 $\mu$G (e.g. Ciolek \& Basu 2000; Levin et al. 2001; Crutcher et al. 2004; Kirk, Ward-Thompson \& Crutcher 2006; Turner \& Heiles 2006). Magnetic pressure becomes comparable to the thermal pressure when the magnetic field strength reaches over 100 $\mu$G. Our stability analysis is not accurate for starless cores with magnetic fields of $\geq$ 100 $\mu$G. Inclusion of magnetic term in the virial theorem is not practical for a dimensionless analysis because the inherent geometry of the field must be taken into account. Treatment of magnetic fields is beyond the scope of this current work.

\subsection{Comparison to Observed Starless Cores with Infall}

We compare the measured infall speeds and dimensionless sizes of observed starless cores which have clear infall signatures and well determined outer radii in Figure \ref{fig6}.  The number of cores used in this study is limited because either only the density or the velocity structure is usually studied from observations of starless cores, but rarely both.  The plotted lines are the same with those in Figure \ref{fig3}. The observed starless cores are marked with black filled circles. The $\xi_{\rm max}$ value of L694-2 is from Harvey et al. (2003b). The peak infall velocity of L694-2 is measured by Lee et al. (2007) using the three hyperfine line of HCN J=1-0. The dimensionless size of L1544 is estimated by Kirk et al. (2005). The infall velocity of L1544 is studied by William, Lee \& Myers (2006) with interferometric observation of N$_2$H$^+$ J=1-0. For L63, we quote the dimensionless size of $\xi_{\rm max}=15$ from Kirk et al. (2005). There is no published infall speed study of L63, therefore,  we estimate the infall velocity by applying the two-layer model (equation (9) of Myers et al 1996) to the HCN lines observed by Sohn et al. (2007). From the HCN $J$=1-0 $F$=1-1 line, we estimate the infall speed of L63 to be about 0.1 km s$^{-1}$. The dimensionless size of L1689B is measured by Dapp \& Basu (2009) and the infall speed is measured by Bacmann \& Pagani (2008).

\begin{figure}[tbh]
  \begin{center}
    \leavevmode
    \epsfxsize = 16.5cm
    \epsfysize = 14.cm
    \epsffile{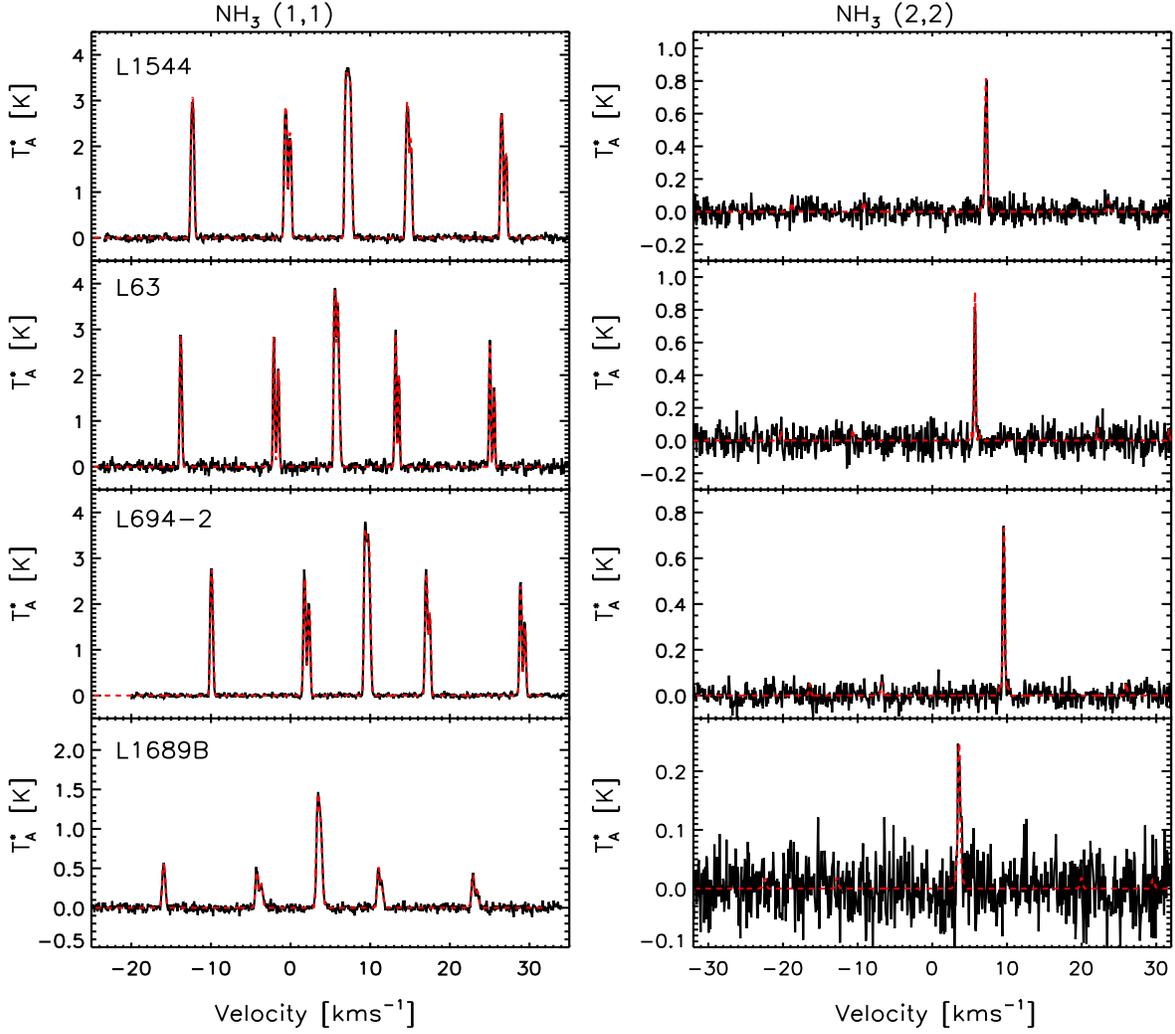}
  \end{center}
\caption{{ \sl Ammonia (1,1) and (2,2) lines observed toward starless cores. The black solid lines are the observed lines and the red dashed lines are fitted models. Fitting is done in the five parameters space including $v_{\rm lsr}$, line width, total optical depth of (1,1) transition, kinetic temperature, and filling factor. The best-fit parameters are listed in Table 2.
}}
\label{fig4}
\end{figure}

\begin{figure}[tbh]
  \begin{center}
    \leavevmode
    \epsfxsize = 16.5cm
    \epsfysize = 14.cm
    \epsffile{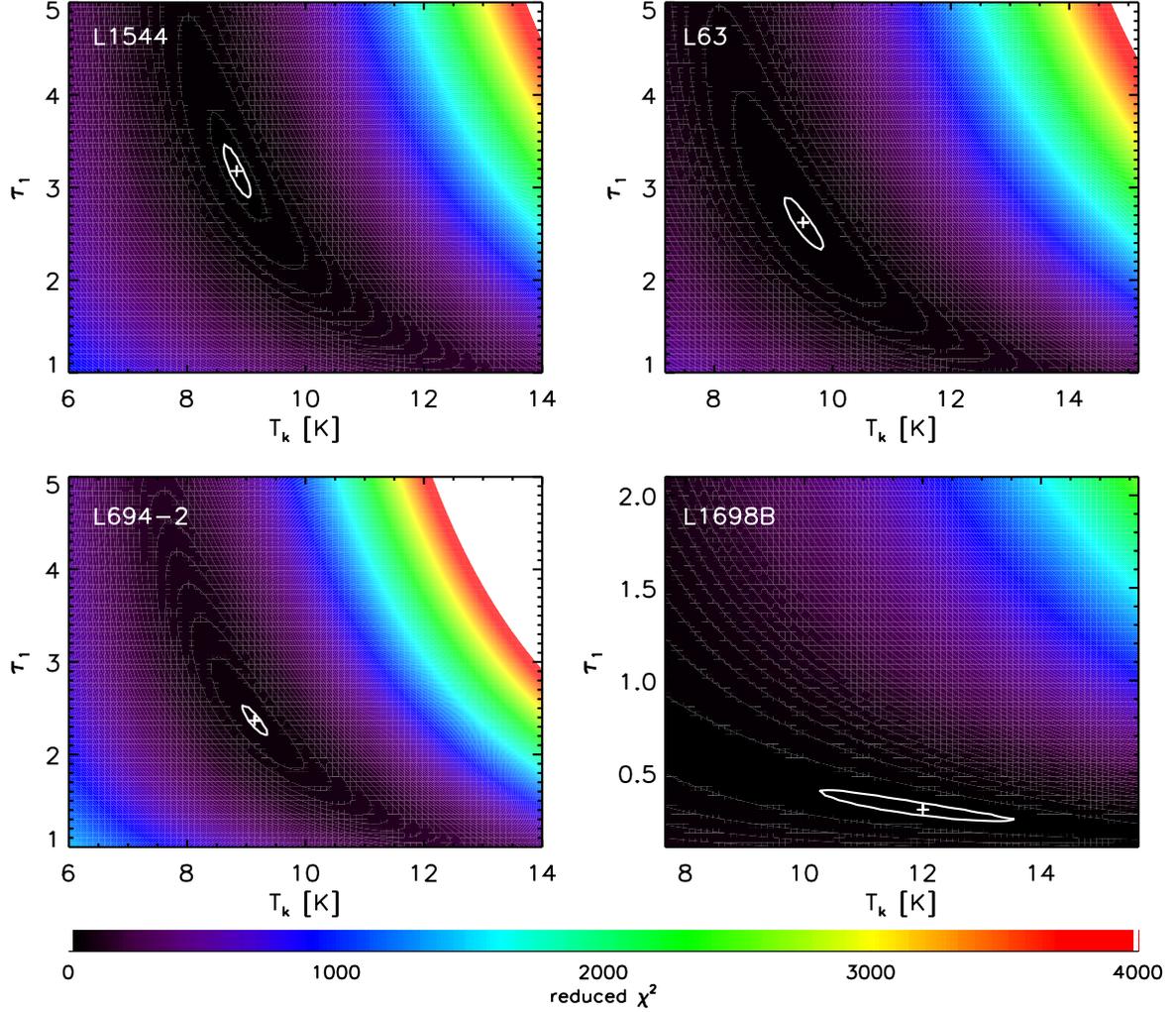}
  \end{center}
\caption{{ \sl Reduced $\chi^2_r$ values in $T_{\rm k}$ $vs.$ $\tau_1$ space. The other three parameters are fixed to be the best-fit values at the shown planes. The white crosses are the positions of the minimum reduced $\chi^2_r$ and the white lines represents the 1$\sigma$ uncertainty space. Uncertainties of each parameters are listed in Table 2.
}}
\label{fig5}
\end{figure}

\begin{figure}[tbh]
  \begin{center}
    \leavevmode
    \epsfxsize = 13.5cm
    \epsfysize = 11.cm
    \epsffile{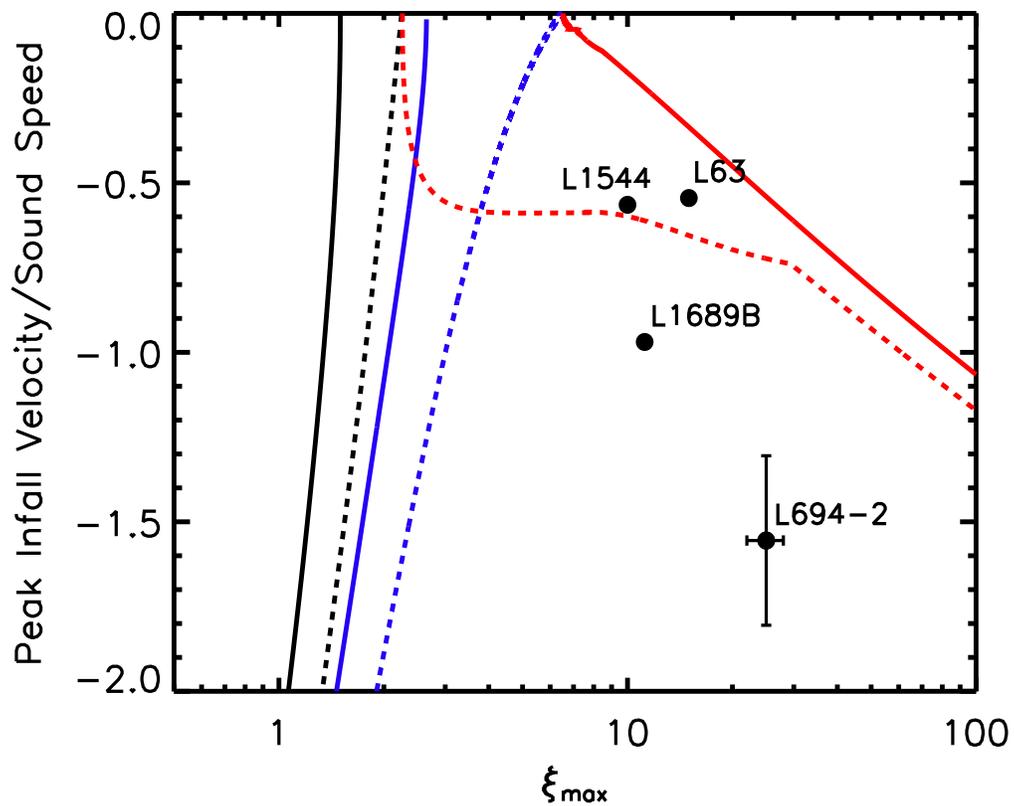}
  \end{center}
\caption{{ \sl Starless cores with clear infall signatures, L63, L1544, L1689B, and L694-2 are plotted with the generalized BE criterion and the limiting cases of the gravitational collapse for Figure 3. All cores are in unstable regime and gravitationally collapsing. The collapse of L694-2 and L1689B seem to be considerably enhanced by external perturbation.
}}
\label{fig6}
\end{figure}

To normalize the infall speeds, internal temperature of those starless cores are required. Since internal temperatures of L63, L694-2 and L1689B are not studied with high precision, we deduced kinetic temperatures of the starless cores utilizing the hyperfine lines of NH$_3$ (1,1) and (2,2) inversion transitions. We fit ammonia spectra with five parameters including $v_{\rm lsr}$, line width $\sigma_{\nu}$, kinetic temperature $T_{\rm k}$, optical depth of (1,1) transition $\tau_1$, and filling fraction $\eta_f$ using the same LTE assumption and method in Rosolowsky et al. (2008). We fit signals with Signal-to-Noise Ratio $> 2$ and estimate the goodness of fit using the reduced $\chi^2_r$. We calculated the reduced $\chi^2_r$ in the full five parameter space instead of using a covariant matrix to more accurately determine 1$\sigma$ uncertainties in the parameters. The results of the fittings are shown in Figure \ref{fig4}. The black solid lines represent observed ammonia lines and red dashed lines represent fitted line profiles. Reduced $\chi^2_r$  at the vicinity of the minimum value are plotted in Figure \ref{fig5}. The cross marks are the minimum points and the white lines represents 1$\sigma$ uncertainty space in $T_{\rm k}$ $vs.$ $\tau_1$ space at the minima of $v_{\rm lsr}$, $\sigma_{\rm v}$, and $\eta_f$ (see Table 2). The temperatures of L1544, L63, L694-2, L1689B from our study are 8.84$\pm$0.30, 9.50$\pm$0.41, 9.14$\pm$0.2, and 12.0$\pm$1.7 K, respectively. Crapsi et al. (2007) also deduced average temperature of L1544 using the same transition of NH$_3$ hyperfine lines, but with interferometric observations, to be 8.75K. Our results are consist within 1$\sigma$ uncertainty. The temperature of L694-2 also agrees with the previous finding of Williams et al. (2006) which is 9.3K. The temperature of L1689B has a relatively large uncertainty. This is because L1689B appears to be optically thin in the NH$_3$ (1,1) line and determining its optical depth is degenerated with filling fraction.

Combining all physical quantities, four starless cores are plotted in Figure \ref{fig6}. We mark error bars only for L694-2 because the quoted references do not provide any error for $\xi_{\rm max}$ and infall speeds for the other three cores. Generally, $\xi_{\rm max}$ has about 50\% uncertainty, and errors of infall speeds are hard to estimate unless a starless core is mapped with high angular resolution and with multiple molecular lines. Accepting that the observed physical quantities for the starless cores quoted are the best determination to date, we may say that all four cores are located in the regime below the collapse of the critical BE sphere. Two cores, L1689B and L694-2, seem to collapse faster than the collapse of the critical uniform density sphere. These two starless cores may be undergoing enhanced collapse by external perturbations, for example, turbulence, or a sudden increase of external pressure. Observations of their surrounding environments are needed to understand the nature of the perturbation. The other two starless cores L1544 and L63 seem to be in quasi-equilibrium collapse.

\section{SUMMARY \& CCONCLUSIONS}

Starless cores are the precursors of low-mass star formation. Their stability and dynamics are important issues in developing a comprehensive picture of how low-mass stars form. However, owing to the difficulties of observing internal motions of starless cores through molecular line observations, only their density structures have been previously considered in investigating their stability and dynamics. In this study we find a stability condition for starless cores with internal motions. We first generalize the conventional stability condition of BE spheres by taking inward homologous motions into account. Even though homologous inward motions may not be a realistic internal motion that describes the observed internal motions of starless cores, it provides a rough estimate how the stability conditions of starless cores vary with the speeds of inward motions. Second, we suggest two limiting cases of spontaneous gravitational collapse: the collapse of the critical BE sphere and the collapse of uniform density spheres. The former model represents the slowest collapse by the self-gravity while the latter model sets the boundary for the fastest infall speed of spontaneous gravitational collapse at a given density. Any starless cores which considerably deviates from these two limiting models are likely to be strongly perturbed by the surrounding environments.

We applied our study to four starless cores which have strong infall signatures (large blue peak asymmetry). Some cores appear to collapse faster than infall speeds permitted by spontaneous gravitational collapse. This result suggests that external perturbations may be important in the collapse dynamics of low-mass starless cores. Further study on the density and velocity structures of starless cores and their surrounding environments are required to obtain more general conclusions on the collapse dynamics of low-mass star formation.

\acknowledgments
We thank anonymous referee for useful comments and corrections.

\newpage
\begin{deluxetable}{ c c c c c c }
  \tablewidth{0pt}
  \tablecaption{Observed Starless Cores}
  \startdata
  \hline
    Source   &  RA (J2000.0)  &     DEC (J2000.0)   &   I(T$_{\rm mb}$) [K km/s]  &    $\sigma_{\rm I}$ [Kkm/s] &   $\sigma_{\rm T_{\rm mb}}$ [K] \\
    L1544    &  05:04:17.2    &     +25:10:43.7     &   14.79                     &      1.01                   &   0.064                         \\
    L63      &  16:50:14.9    &     -18:06:22.5     &   10.22                     &      0.72                   &   0.107                         \\
    L694-2   &  19:41:04.3    &     +10:57:00.7     &   12.13                     &      0.83                   &   0.078                         \\
    L1689B   &  16:34:48.3    &     -24:38:03.5     &   2.962                     &      0.26                   &   0.093                         \\
  \enddata
\end{deluxetable}
\begin{deluxetable}{ c c c c c c }
  \tablewidth{0pt}
  \tablecaption{Physical Quantities of Starless Cores Deduced from NH$_3$ (1,1) and (2,2)}
  \startdata
  \hline
  Source & v$_{\rm lsr}^a$ [kms$^{-1}$] & $\sigma_\nu$ [kms$^{-1}$] & T$_{\rm k}$ [k] & $ \tau_1 $ & $\eta_f$   \\
  \hline
   L1544   & 7.196  & 0.134$_{-0.005}^{+0.005}$    &  8.84$_{-0.30}^{+0.30}$   &   3.17$_{-0.35}^{+0.41}$   &  0.80$_{-0.05}^{+0.06}$    \\
   L63     & 5.721  & 0.0825$_{-0.0026}^{+0.0053}$ &  9.50$_{-0.41}^{+0.41}$   &   2.62$_{-0.35}^{+0.35}$   &  0.80$_{-0.06}^{+0.06}$    \\
   L694-2  & 9.595  & 0.127$_{-0.003}^{+0.003}$    &  9.14$_{-0.20}^{+0.20}$   &   2.36$_{-0.15}^{+0.15}$   &  0.77$_{-0.04}^{+0.04}$    \\
   L1689B  & 3.591  & 0.172$_{-0.032}^{+0.018}$    &  12.0$_{-1.7}^{+1.7}$     &   0.30$_{-0.20}^{+0.20}$   &  0.47$_{-0.21}^{+0.52}$    \\
  \enddata
  \tablenotetext{a}{\footnotesize Uncertainties of v$_{\rm lsr}$ are not estimated}
\end{deluxetable}

\newpage

\newpage

\newpage

\newpage

%%%%%%%%%%%%%%%%%%%%%%%%%%%%%%%%%%%%%%%%%%%%%%%%%%%%%%%%%%%%%%%%%%%%%%%%%%%%%%%%%%%%%%%%%%%%%%%%%%%%%%%%%%%%%%%%%%%%%%%%%%%%%%%%%%%%%%%
%%%%%%%%%%%%%%%%%%%%%%%%%%%%%%%%%%%%%%%%%%%%%%%%%%%%%%%%%%%%%%%%%%%%%%%%%%%%%%%%%%%%%%%%%%%%%%%%%%%%%%%%%%%%%%%%%%%%%%%%%%%%%%%%%%%%%%%
%%%%%%%%%%%%%%%%%%%%%%%%%%%%%%%%%%%%%%%%%%%%%%%%%%%%%%%%%%%%%%%%%%%%%%%%%%%%%%%%%%%%%%%%%%%%%%%%%%%%%%%%%%%%%%%%%%%%%%%%%%%%%%%%%%%%%%%
%%%%%%%%%%%%%%%%%%%%%%%%%%%%%%%%%%%%%%%%%%%%%%%%%%%%%%%%%%%%%%%%%%%%%%%%%%%%%%%%%%%%%%%%%%%%%%%%%%%%%%%%%%%%%%%%%%%%%%%%%%%%%%%%%%%%%%%
%%%%%%%%%%%%%%%%%%%%%%%%%%%%%%%%%%%%%%%%%%%%%%%%%%%%%%%%%%%%%%%%%%%%%%%%%%%%%%%%%%%%%%%%%%%%%%%%%%%%%%%%%%%%%%%%%%%%%%%%%%%%%%%%%%%%%%%
%%%%%%%%%%%%%%%%%%%%%%%%%%%%%%%%%%%%%%%%%%%%%%%%%%%%%%%%%%%%%%%%%%%%%%%%%%%%%%%%%%%%%%%%%%%%%%%%%%%%%%%%%%%%%%%%%%%%%%%%%%%%%%%%%%%%%%%
%%%%%%%%%%%%%%%%%%%%%%%%%%%%%%%%%%%%%%%%%%%%%%%%%%%%%%%%%%%%%%%%%%%%%%%%%%%%%%%%%%%%%%%%%%%%%%%%%%%%%%%%%%%%%%%%%%%%%%%%%%%%%%%%%%%%%%%
%%%%%%%%%%%%%%%%%%%%%%%%%%%%%%%%%%%%%%%%%%%%%%%%%%%%%%%%%%%%%%%%%%%%%%%%%%%%%%%%%%%%%%%%%%%%%%%%%%%%%%%%%%%%%%%%%%%%%%%%%%%%%%%%%%%%%%%
%%%%%%%%%%%%%%%%%%%%%%%%%%%%%%%%%%%%%%%%%%%%%%%%%%%%%%%%%%%%%%%%%%%%%%%%%%%%%%%%%%%%%%%%%%%%%%%%%%%%%%%%%%%%%%%%%%%%%%%%%%%%%%%%%%%%%%%
%%%%%%%%%%%%%%%%%%%%%%%%%%%%%%%%%%%%%%%%%%%%%%%%%%%%%%%%%%%%%%%%%%%%%%%%%%%%%%%%%%%%%%%%%%%%%%%%%%%%%%%%%%%%%%%%%%%%%%%%%%%%%%%%%%%%%%%

\end{document}